\documentstyle[jltp,epsfig,11pt,twoside]{article}

\title{Supercurrent in a mesoscopic proximity wire}

\author{Frank K. Wilhelm, Andrei D. Zaikin, and Gerd Sch\"{o}n
\address{Institut f\"{u}r Theoretische Festk\"{o}rperphysik,
Universit\"{a}t Karlsruhe (TH), D-76128 Karlsruhe, Germany}}
\runninghead{Frank K. Wilhelm, Andrei D. Zaikin, and Gerd
Sch\"{o}n}{Superconducting properties of thin proximity wires}

\begin{document}
\begin{abstract}
{Recent experiments on the proximity induced supercurrent in
mesoscopic normal wires revealed a surprising temperature dependence. 
They suggest clean-limit behavior although the wires are
strongly disordered.
We demonstrate that this unexpected scaling is actually contained in 
the conventional description of diffusive superconductors
and find excellent agreement with the experimental results. 
In addition we propose a SQUID-like proximity structure for
further experimental investigations of the effects in question.}

\end{abstract}

\maketitle

\section{Introduction}

A normal metal in
direct contact with a superconductor acquires superconducting
properties \cite{deG,Parks:deG}. Although this proximity effect 
has been discussed for a long time, it has  recently attracted
new attention because of the dramatic progress in nanotechnology
which allows the fabrication and study of metallic structures
in the mesoscopic regime. 
Due to the proximity effect a supercurrent can flow through a normal
metal between two superconductors \cite{Kulik,Likharev,Zaikin}. 
Recently, this current has been detected by Courtois et al. \cite{Cour} 
in a thin normal wire in the diffusive regime with superconducting strips
deposited on its top (see Fig. 1). 

\begin{figure}
\centerline{\psfig{figure=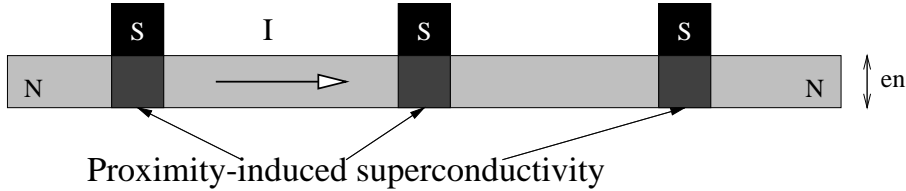,width=120mm}}
\caption{Experimental geometry}
\end{figure}

The supercurrent is determined by
an overlap of Cooper-pair wave functions penetrating into
the normal wire from the superconducting strips. Since in the diffusive
limit the effective penetration length is of order \cite{deG,Parks:deG} 
$ \xi_N=\sqrt{D/2\pi T}$ one should expect that the supercurrent 
in the system is proportional to \cite{Likharev,Zaikin} 
$I_c(T)\propto e^{-\sqrt{T/T_0}}$. In contrast, 
the experimental data of Ref. \cite{Cour} showed a much
better fit to the dependence $I_c(T)\propto e^{-{T/T_1}}$, which is
typical for ballistic systems \cite{Kulik}. Furthermore, the fitting parameter
$T_1$ deviated from standard clean-limit results by more than one
order of magnitude.
This has raised the question,
whether the usual criterion distinguishing between 
ballistic and diffusive limits is correct when
applied to proximity induced superconductivity in the normal metal,
or whether the above phenomenon has to be attributed
to quantum effects not contained in the quasiclassical theory of
superconductivity. 

Below we will demonstrate that neither of these conjectures is true. 
We employ a quasiclassical calculation of the supercurrent in the 
proximity structures of Fig. 1 and show the corresponding results 
agree well with the experimental data \cite{Cour}. 
Furthermore, we will also propose a new, equivalent experiment, 
where superconducting material is deposited onto a normal
ring. In this 'proximity SQUID' the magnetic flux through the ring
is the equivalent of the phase
difference across the normal wire.

\section{The Model and the Formalism}

Since the critical current of the structure of Fig.1 is determined 
by the longest SNS-cell which
serves as a ``bottleneck'' it is sufficient to study a single cell.
We denote the distance between these two adjacent superconducting
strips as $d$ and consider the case where $d$ is much larger than the
strips' thickness.

We will use the standard formalism of quasiclassical Green's
functions \cite{Eilen} in the diffusive limit
\cite{Usa}

For a thin normal wire with a
thickness $\ll \xi_N$ (this condition is well justified in the experiment
\cite{Cour}) the Green functions in the wire directly below
superconducting strips are equal to those of a superconductor
for all relevant energies  \cite{Kup}.
In what follows we further assume that the superconductors have bulk
properties, neglecting any suppression of the superconducting
gap $\Delta$ in the strips in the vicinity of SN boundaries. 

\section{Solution and Results}

Details of the formalsim and the calculation will be given in a
future publication, so we will only present the key results here.

A characteristic energy scale is provided by the Thouless energy
$E_d=D/d^2$, where $D$ is the diffusion constant.
For high temperatures, $T\gg E_d$ (equivalent to the
geometrical condition $d\gg\xi_N$), the mutual influence between the
superconductors can be neglected.
In this approximation, we get the current-phase relation
$I = I_c \sin(\varphi)$ with critical current
\begin{equation}
	I_c={64\pi\over{3+2\sqrt{2}}}{T\over eR_N}{d\over\xi_N}
	\exp\left({-{d\over\xi_N}}\right)\propto T^q\exp\left(-\sqrt{T\over T_0}\right)
\label{simpstrom}
\end{equation}
with $2\pi T_0=E_d$, $q=3/2$ and $R_N$ is the normal state resistance of
the normal part. For comparison,
the expression following from the Ginzburg-Landau analysis 
\cite{Parks:deG} has the same $T$-dependence but with $q=1/2$.
On the other hand, in the clean limit one expects a temperature
dependence  $I_c\propto e^{-T/T_1}$. 

At this stage we note  a remarkable
mathematical artefact caused by the exponent $q=3/2$. 
The logarithmical derivative of (\ref{simpstrom}) 
$${dI_c(T)\over
I_c(T)dT}=\left({3\over2T}-{1\over2\sqrt{TT_0}}\right)$$
 has a
minimum at $T = 36 T_0$ and varies very slowly at higher temperatures, so
$\log I_c$ is almost linear in $T$. 
As a good approximation for the slope in a logplot, we can
take the logarithmical derivative in the minimum and get
 $I_c\propto e^{-T/T^\ast}$ where
$T^\ast=24T_0$. 
This implies that  within a considerable temperature interval a
``quasi-clean'' scaling is found also in dirty SNS systems,  
which explains the behavior
found in the experiments of Courtois et al. \cite{Cour}.

In the low temperature limit $T=0$ another simple estimate for the 
critical current can be derived.
\begin{equation}
	I_c={\Delta\over
	R_Ne}\arctan\left({E_d\over2\Delta}\right)
	\stackrel{E_d\ll\Delta}{\longrightarrow}{D\over2R_Ned^2}\;.
\label{Ic}
\end{equation}
This demonstrates the importance of the Thouless energy
as a relevant energy scale in SNS-junctions at $E_d\ll\Delta$. 
The Thouless energy also determines
the proximity induced effective gap in the quasiparticle spectrum of 
the N-metal (see Ref. \cite{GWZ} and further references
 therein). We further note that  (\ref{Ic}) resembles
 the well-known Ambegaokar-Baratoff formula for
the critical current of a Josephson tunnel junction at $T=0$
 if we substitute $D/d^2$  by the gap $\sim \Delta$. Thus our results
emphasize that the Thouless energy in a diffusive proximity coupled 
normal wire plays the same role as the gap $\Delta$ in 
a ``strong'' superconductor and does not only determine the density of
states, but also the critical current.

\begin{figure}
\centerline{\psfig{figure=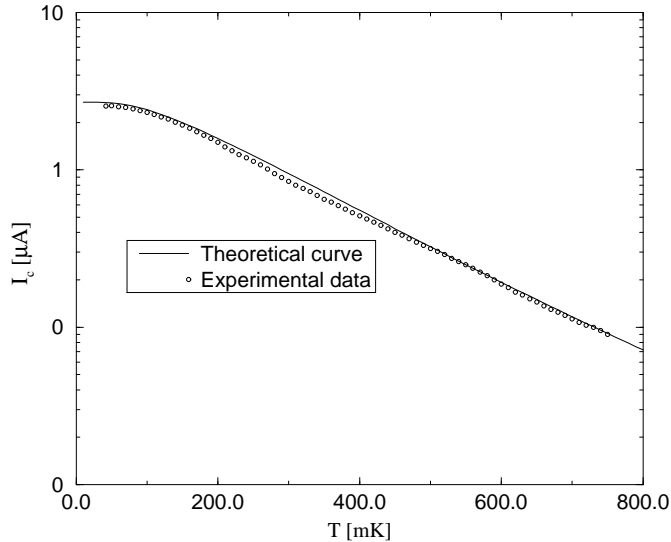,width=10cm}}
\caption{Critical current for the experimental parameters [6],
experimental data are supplied by H.Courtois}
\end{figure}

A solution of the full problem as
obtained numerically is shown in fig 2.
The result matches
quantitatively to the experiments, if $d$ is chosen
 slightly larger than the average cell length of the experiment. 
This reflects the
fact, that the critical current of the whole chain is determined by
the longest SNS-cell with the lowest $I_c$. 
The adjustment lies within the range of experimental accuracy.
 In general the current-phase relation 
$I(\varphi)$  may deviate from sinusodiality \cite{Likharev,Zaikin}. 
We investigated also this property, but found in the SNS geometry
nearly no deviation down to very low temperatures (see Fig. 3).

\begin{figure}
\centerline{\psfig{figure=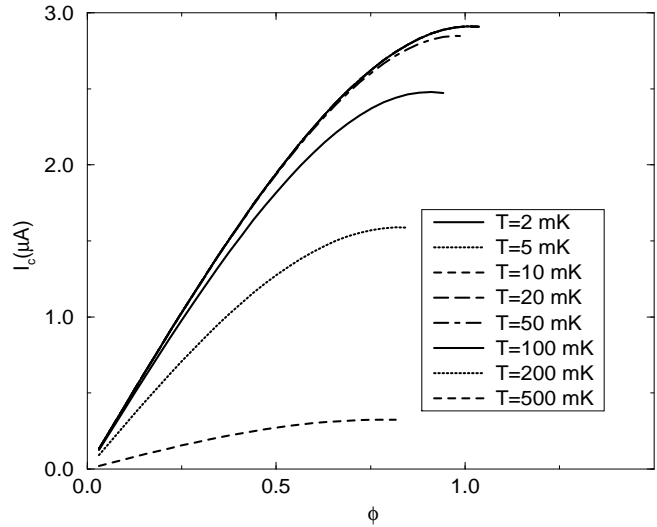,width=10cm,height=8cm}}
\caption{Current phase relation, parameters as in the experiment, see
Fig. 2}
\end{figure}

\section{The Proximity SQUID}

The proximity induced supercurrent should lead to
an interesting flux-periodic behavior of the ``proximity loop''
structure shown in Fig. 4. A normal ring  is contacted over a range of
lentgh $d_S$ by superconducting material.
If the ring is narrow it can be mapped onto the linear system
discussed above by absorbing the vector potential in the gauge 
invariant phase 
$\varphi(r)=\varphi_0(r)+2e\int_0^r dr^\prime\;A(r^\prime)$.
I.e. the anomalous
Green's function $F$ carries the phase factor $\exp[i\varphi_0(r)]$, but
$\alpha(r)$ and $\varphi(r)$ satisfy the Usadel equation
with appropriate boundary conditions and the substitution $d=L-d_S$. 
Since the Green's functions must be single-valued at
every point of the ring, the ``real'' phase $\varphi_0$ can
change only by multiples of $2\pi$ when circling around the ring.
For this reason, at $x=0$ the gauge-invariant phase drops by
$$\phi=\varphi(0+)-\varphi(0-)=2e\oint dr\;A(r)={2e\Phi\over\hbar},$$
where $\Phi$ is
the magnetic flux through the ring. 

\begin{figure}
\centering
{\psfig{figure=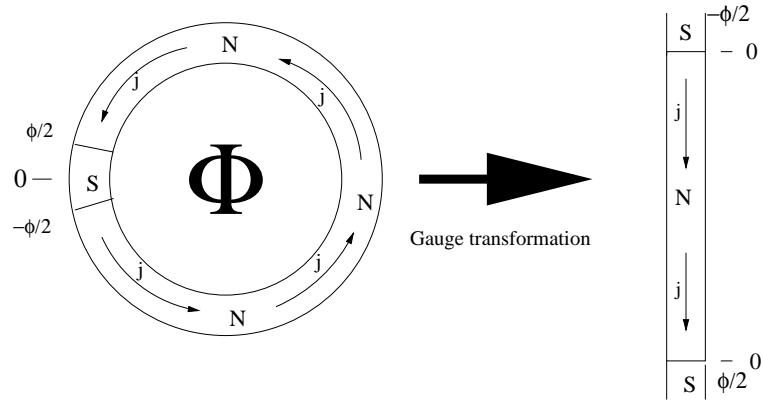,width=10cm}}
\caption{Proximity SQUID}
\end{figure}

Taking into account the self-inductance, this system is
completely equivalent to
a standard SQUID formed by a superconducting loop interrupted
by a weak link. The configuration considered here is 
complementary: a normal loop is interrupted by a narrow superconducting
strip, but due to the proximity effect it shows the
same properties as the standard SQUID. 

At relatively high temperatures the supercurrent in the normal proximity
loop is exponentially small. However, as the temperature is lowered below
the corresponding Thouless energy the supercurrent becomes 
large (cf. (\ref{Ic})) and
can be easily detected experimentally. Also the current-phase relation
of a diffusive SNS structure can be easily studied.

\section{Conclusions}
Making use of a standard quasiclassical formalism of the superconductivity
theory we evaluated the supercurrent in mesoscopic proximity wires.
Our results match quantitatively with the experimental data \cite{Cour}
thus demonstrating that the quasiclassical theory of superconductivity
is sufficient for the description of these systems.
The deviations of the
current-phase relation from sinusodiality are found to be small even
at very low temperatures. Finally we argue that a normal-metall loop
interrupted by a narrow superconducting strip has the same properties 
as a standard SQUID and can be used for further experimental investigations 
of the proximity effect in mesoscopic systems.

We would like to thank C. Bruder, H. Courtois and B. Pannetier for
useful discussions. This work was supported by the Deutsche
Forschungsgemeinschaft through Sonderforschungsbereich 195. One of us
(G.S.) acknowledges the A.v. Humboldt award of the Academy of Finland.
and the hospitality of the Helsinki University of Technology.

\end{document}